\documentclass[prd,twocolumn,floatfix,preprintnumbers,nofootinbib]{revtex4-2}

\usepackage[utf8]{inputenc}
\usepackage{subfigure}
\usepackage[normalem]{ulem}
\usepackage[dvipsnames]{xcolor}
\usepackage{mathrsfs}
\usepackage{amssymb}
\usepackage{amsmath}
\usepackage{mathtools}
\usepackage{graphicx}
\usepackage{physics}
\usepackage{accents}
\usepackage{comment}
\usepackage{placeins}

\usepackage{tabularx}

\newcommand{\ud}{\mathrm{d}}
\renewcommand{\tr}{\mathrm{Tr}}

\newcommand{\nc}{N_\mathrm{c}}
\newcommand{\hati}{\hat{\imath}}
\newcommand{\hatj}{\hat{\jmath}}

\definecolor{lcolor}{rgb}{0.5,0,0}
\definecolor{citcolor}{rgb}{0,0.3,0.0}

\usepackage[breaklinks,colorlinks,citecolor=citcolor,urlcolor=blue,linkcolor=lcolor]{hyperref}
\usepackage[capitalise]{cleveref}

\begin{document}

\author{Pooja}
 \email{pooja.p.pooja@jyu.fi}
\author{Dana Avramescu}
\email{dana.d.avramescu@jyu.fi}
\author{Tuomas Lappi}
\email{tuomas.v.v.lappi@jyu.fi}
\affiliation{Department of Physics, University of Jyväskylä,  P.O. Box 35, 40014 University of Jyväskylä, Finland}
\affiliation{Helsinki Institute of Physics, P.O. Box 64, 00014 University of Helsinki, Finland}

\title{The leading Lyapunov exponent in the glasma}

\begin{abstract}
We show that small perturbations in the boost-invariant color fields of the glasma exhibit an exponential growth with the square root of time. We interpret this growth rate as a Lyapunov exponent, related to entropy production and the thermalization timescale in the earliest stage of heavy-ion collisions.  Working in a regime that is linear in this perturbation, we extract the time dependence of this mode as $\sim \exp(0.4\sqrt{g^2\mu\tau})$ for SU($2$), where $g^2\mu$ is proportional to the saturation scale and the square-root dependence is caused by the boost-invariant expansion of the system. We show that the growth rate of this mode is, unlike its amplitude, remarkably insensitive to the details of how the perturbations are initialized. In particular, we show that the unstable mode couples to all momentum scales present in the initial perturbation.
\end{abstract}

\maketitle 


\section{Introduction}
\label{Section: Introduction}

The initial stage of a heavy-ion collision is believed to be dominated by gluons, in a very strongly overoccupied, far-from-equilibrium state referred to as the glasma~\cite{Lappi:2006fp, Gelis:2012ri}. The large overoccupation permits, in the weak-coupling limit, a description of the system in terms of classical gauge fields. This system then evolves in time towards a quark-gluon plasma in a thermalization process~\cite{Berges:2020fwq} that is still poorly understood. 

The focus of much of the effort to understand thermalization in the early stages of heavy-ion collisions has been on \emph{isotropization}, namely the process of going from the very anisotropic energy-momentum tensor in the earliest stage towards isotropy in the local rest frame of the plasma, connecting to a description in terms of hydrodynamics. While isotropization is important for phenomenology, there are also more fundamental aspects to the question of thermalization, for example, in terms of the relation between quantum and classical physics, entanglement, and chaoticity~\cite{Kharzeev:2026jkq}. 

Classical chaoticity is seen in the exponential growth of small perturbations. It is most straightforwardly quantified in terms of the growth rates, i.e., Lyapunov exponents. Chaotic behavior has been studied in classical Yang-Mills (CYM) systems in thermal equilibrium already some time ago~\cite{Muller:1992iw, Biro:1993qc, Heinz:1996wx, Kunihiro:2010tg, Bazak:2023kol}.  The applicability of CYM to thermal systems is, however, limited by the Rayleigh-Jeans catastrophe: the degrees of freedom carrying most of the energy in a thermal system are not classical. However, the situation in the glasma is different: the overoccupation of the dominant momentum modes does make the behavior of the system tractable with classical fields. Indeed, the unstable behavior related to isotropization has been studied with CYM methods~\cite{Romatschke:2005pm, Romatschke:2006nk, Epelbaum:2013ekf}, and Lyapunov exponents have been extracted for non-equilibrium classical field configurations~\cite{Iida:2013qwa}. Looking at chaotic behavior in the glasma is thus a promising avenue to understand thermalization and entropy production in the glasma~\cite{Tsukiji:2017pjx}, and to understand thermalization dynamics in overoccupied systems more generally~\cite{Muller:2024rjh, Guin:2025lpy}.

In this paper, our goal is to demonstrate the existence of chaotic, exponentially growing modes in the boost-invariant 2+1-dimensional glasma fields. We will then extract a numerical value for the Lyapunov exponent.  Rather than going into much detail about the definition of entropy or chaos in a classical statistical system, we will just posit that the growth rate provides a timescale of entropy production and thermalization in the glasma.  We will study a ``standard'' glasma setup (although in this first study with SU($2$)) using the McLerran-Venugopalan (MV) model~\cite{McLerran:1993ni, McLerran:1993ka, McLerran:1994vd}, where we have a rough understanding of what the physically realistic magnitude of the color charge density parameter $g^2\mu$ is. While we will express our results for the time dependence in terms of the dimensionless scaling variable $g^2\mu \tau$, an estimate in units of fm$/c$ could be obtained by taking $g^2\mu$ to be of the order of the saturation scale $Q_\mathrm{s}$ (see Ref.~\cite{Lappi:2007ku}), of order $1\dots 2$ GeV at the LHC (up to an uncertainty of a factor $\lesssim 2$ from the difference between SU($2$) and SU($3$)). 

We will approach the problem here with a very pedestrian, but robust, method. We will initialize infinitesimally small fluctuations in the classical color fields at the initial time of the collision. We will then follow the time evolution of these fluctuations until we can identify a clear exponential growth, while verifying that we still stay in a regime where the dynamics is linear in the amplitude of the fluctuations.  Because of the boost-invariant expansion of the glasma, we expect the growth to not be exponential in time, but exponential in the square root of the proper time $\tau$. By fitting this time dependence with a specific functional form, we will extract the value of the leading Lyapunov exponent, i.e., the one that dominates the late-time behavior. Extracting the whole spectrum of Lyapunov exponents and analyzing the structure of the corresponding fluctuations is left to future work. We will, however, gain some indication of the structure of the leading exponential mode by varying the momentum-space structure of the fluctuations at the initial time. 

This paper is structured as follows. 
We will first briefly review how the boost-invariant glasma fields in the MV model are obtained, both in the continuum and on the lattice, in Sec.~\ref{Section: Framework_Glasma}. Then, in Sec.~\ref{Section: Formalism_Perturbation}, we will describe in detail how we initialize the fluctuations in the glasma fields. In Sec.~\ref{Section: Lyapunov_Extraction}, we show our first result for the time dependence of the fluctuations and use it to develop our fitting methodology to extract the Lyapunov exponents. Using this method, we then, in Sec.~\ref{Section: Results}, analyze a broader range of initial conditions for the fluctuations, before concluding in Sec.~\ref{Section: Conclusions}.

\section{Theoretical Framework}
\label{Section: Framework_Glasma}
\subsection{Glasma fields}

The most convenient way to quantitatively describe the overoccupied gluonic system in the initial stage of heavy-ion collisions is to use the Color Glass Condensate (CGC) effective theory. Here, the fast, large-$x$ partons of the colliding nuclei act as static color sources for the soft gluon fields representing the small-$x$ gluons. When these interact in the collision, they first form an overoccupied gluonic system known as the glasma, and subsequently evolve towards an equilibrated quark-gluon plasma.

In this work, we generate the initial conditions for the classical gluon fields within the framework of the MV model~\cite{McLerran:1993ni, McLerran:1993ka, McLerran:1994vd}. In this approach, the large-$x$ partons are modeled as randomly distributed color charges on a two-dimensional plane, representing the Lorentz-contracted nucleus. The distribution of these color charges, $\rho^a(\mathbf{x}_\perp)$, within the two colliding nuclei, labeled as $L$ (left-moving) and $R$ (right-moving), is taken to be a local Gaussian ensemble. This is characterized by a vanishing expectation value
\begin{equation}
\langle\rho^a_{L,R}(\mathbf{x}_\perp) \rangle = 0,
\end{equation}
and a local correlation in the transverse plane
\begin{equation}
	\langle \rho^a_{L,R}(\mathbf{x}_\perp)~ \!   \rho^b_{L,R}(\mathbf{y}_\perp) \rangle = g^2 \mu_{L,R}^2 \delta^{ab} \delta^{(2)}(\mathbf{x}_\perp - \mathbf{y}_\perp).
\end{equation}
Here, $a$ and $b$ are adjoint color indices. For $\mathrm{SU}(N_c)$ QCD theory, these indices run from $1$ to $N_c^2-1$. The MV model parameter $\mu$ in the variance represents the color charge density scale and has dimensions of mass, and $g$ is the coupling constant. Throughout this paper, the subscript $\perp$ indicates quantities in the transverse plane.

Once the color charge density has been generated for each nucleus, the next step is to determine the corresponding initial gauge fields $A^\mu$ before the collision. The interaction between the hard, static color charges  and the soft, dynamical gauge fields is governed by the Yang-Mills equations
\begin{equation}
	D_\mu F^{\mu\nu} = J^\nu,
\label{Eq: CYM_with_source}
\end{equation}
where $J^\nu$ denotes the color current of the fast partons composing the nucleus. The covariant derivative is defined as $D_\mu(\cdot) = \partial_\mu(\cdot) + ig[A_\mu,\,\cdot\,]$ and the non-Abelian field strength tensor is
\begin{equation}
	F^{\mu\nu}	= \partial^\mu A^\nu - \partial^\nu A^\mu	+ ig[A^\mu, A^\nu].
\end{equation}

To solve Eq.~\eqref{Eq: CYM_with_source} for a single nucleus, a specific gauge choice must be made. A convenient starting point is to start with the ``projectile'' light-cone gauge for colliding nuclei, defined by $A^- = 0$ for the right-moving nucleus and $A^+ = 0$ for the left-moving one. At high energy, the evolution in the respective light-cone times ($x^+$ for the right-moving and $x^-$ for the left-moving nucleus) is so slow that it can be neglected;  this is the ``glass'' in the CGC. In this case, the (``projectile'') light-cone gauge solutions to the equations of motion for the single nucleus coincide with the covariant-gauge ones. In this gauge, for a nucleus localized at $x^\mp=0$ (for a right- or left-moving nucleus, respectively), the transverse components of the gauge field of the single nucleus field can be shown to vanish:
\begin{equation}
	A^i = 0, \qquad i = x,y.
\end{equation}
Consequently, the only non-vanishing components of the gauge fields for the single nuclei are 
\begin{equation}\label{eq:singlenuclei}
	A^+_R \equiv \alpha_R(\mathbf{x}_\perp), \quad A^-_L \equiv \alpha_L(\mathbf{x}_\perp).
\end{equation}
In this gauge, the problem  of a single nucleus equation of motion reduces to a two-dimensional Poisson equation in the transverse plane for the fields $\alpha(\mathbf{x}_\perp)$, namely,
\begin{equation}
	\partial_\perp^2 \alpha_{L,R}(\mathbf{x}_\perp) =  -\rho_{L,R}(\mathbf{x}_\perp),
\label{Eq: Poisson_equation_position_space}
\end{equation}
where $\partial_\perp^2$ is the Laplacian in the transverse plane only. This equation directly relates the gauge field $\alpha$ to the color charge density $\rho$ that sources it.

To solve Eq.~\eqref{Eq: Poisson_equation_position_space}, it is convenient to work in momentum space. Fourier transforming both sides gives an algebraic equation for the Fourier modes
\begin{equation}
	\mathbf{k}_\perp^2\,\tilde{\alpha}_{L,R}(\mathbf{k}_\perp)
	= \tilde{\rho}_{L,R}(\mathbf{k}_\perp),
		\label{Eq:Poisson_Eq_solution_momentum_space}
\end{equation}
where $\tilde{\alpha}_{L, R}(\mathbf{k}_\perp)$ and $\tilde{\rho}_{L, R}(\mathbf{k}_\perp)$ denote the Fourier components of the color potential and charge density associated with the corresponding nucleus. These transforms are implemented numerically using the fast Fourier transform (FFT) algorithm. The solution for the gauge field in momentum space is then trivial, i.e.,
\begin{equation}
	\tilde{\alpha}_{L,R}(\mathbf{k}_\perp) = \frac{\tilde{\rho}_{L,R}(\mathbf{k}_\perp)}{\mathbf{k}_\perp^2}.
	\label{Eq: Poisson_Eq_solution_momentum_space_2}
\end{equation}
The zero transverse momentum limit of Eq.~\eqref{Eq: Poisson_Eq_solution_momentum_space_2} needs to be regularized in some way. Often, this is done with a screening mass, replacing $\mathbf{k}_\perp^2 \to \mathbf{k}_\perp^2+ m^2$. However, here we will follow the procedure used in earlier numerical calculations (e.g., \cite{Krasnitz:1999wc, Krasnitz:2000gz, Krasnitz:2001qu, Lappi:2003bi, Krasnitz:2003jw}) and regularize the Poisson equation by leaving out the zero-momentum mode on a discretized lattice. This effectively means that we let the lattice volume act as the infrared regulating scale. Finally, an inverse Fourier transform of $\tilde{\alpha}(\mathbf{k}_{\perp})$ brings the solution back to coordinate space, yielding $\alpha(\mathbf{x}_{\perp})$, the gauge potential that serves as the initial field configuration in the MV model for each nucleus.

In order to treat the collision of two nuclei traveling in opposite directions, we now need to change from the single-nucleus solutions in Eq.~\eqref{eq:singlenuclei} in covariant-gauge to a gauge that can be used to simultaneously describe both of them. This can be done by transforming the fields of both nuclei from their respective projectile light-cone gauges ($A^-_R=0, A^+_L=0$) to their respective ``target'' light-cone gauges ($A^+_R=0, A^-_L=0$). These gauge transformations are performed using the corresponding Wilson lines
\begin{equation}
	V_{L,R}(\mathbf{x}_\perp) = \exp\!\left[\, ig\, \alpha_{L,R}(\mathbf{x}_\perp) \right].
\end{equation}
The new gauge condition makes the dominant covariant gauge fields in Eq.~\eqref{eq:singlenuclei} vanish. Instead, the information on the gluon fields is transferred to the transverse components, which are given by
\begin{equation}
	\alpha^i_{L,R}(\mathbf{x}_\perp) = 	\frac{i}{g}\, 	V_{L,R} (\mathbf{x}_\perp)\, ~ \! \!  \partial^i 	V_{L,R}^\dagger(\mathbf{x}_\perp),	
\end{equation}
where $A_{L,R}^i\equiv\alpha^i_{L,R}(\mathbf{x}_\perp)$ with $i = x,y$.

The glasma gauge fields are defined inside the future light-cone formed after the collision of two ultrarelativistic nuclei. In this limit, the system is effectively invariant under longitudinal Lorentz boosts, and thus any physical observable must respect boost invariance. A natural choice of coordinates for this case is provided by the Milne coordinate system $(\tau,\eta)$, defined by
\begin{equation}
	\tau = \sqrt{2 x^{+} x^{-}}, 	\qquad
	\eta = \frac{1}{2}\ln\!\left(\frac{x^{+}}{x^{-}}\right),
	\label{Eq: MilneDef}
\end{equation}
where $\tau$ is the proper time and $\eta$ the space-time rapidity. A natural gauge condition to use in this problem is the temporal gauge
\begin{equation}
	A^{\tau} = 0.
\end{equation}
We also assume boost invariance of the produced glasma, so that the gauge potentials depend only on $\tau$ and the transverse coordinates,
\begin{equation}
	A^{\mu}(\tau,\eta,\mathbf{x}_{\perp}) 
	= A^{\mu}(\tau,\mathbf{x}_{\perp}).
\end{equation}
Along the boundary of the future light-cone, the non-Abelian superposition of the colliding nuclei gives the glasma initial conditions~\cite{Kovner:1995ts}
\begin{equation}
	A^{i}(\tau,\mathbf{x}_{\perp})
	\bigg|_{\tau=0}
	= \alpha^{i}_{L}(\mathbf{x}_{\perp})
	+ \alpha^{i}_{R}(\mathbf{x}_{\perp}),
	\label{eq:AiIC}
\end{equation}
and the longitudinal component is generated through the commutator of the two transverse pure gauges:
\begin{equation}
	A^{\eta}(\tau,\mathbf{x}_{\perp})
	\bigg|_{\tau=0}
	= \frac{ig}{2}
	\left[
	\alpha^{i}_{L}(\mathbf{x}_{\perp}),\,
	\alpha^{i}_{R}(\mathbf{x}_{\perp})
	\right].
	\label{eq:AetaIC}
\end{equation}
The time derivatives of the fields at $\tau=0$ are zero
\begin{equation}
    \partial_{\tau} A^{i}\big|_{\tau=0} = 0,\quad \partial_{\tau} A^{\eta}\big|_{\tau=0} = 0. 
    \label{eq:StaticIC}
\end{equation}

In practice, the equations of motion are solved in a Hamiltonian formalism, in terms of the fields and their canonical conjugate momenta (electric fields). In the continuum formulation, our canonical coordinates are the gauge potentials $ A_i$ and $A_\eta$. The corresponding canonical momenta are
\begin{equation}
    E^{i} = - \tau\,\partial_{\tau} A^{i},\quad E_{\eta} = \frac{1}{\tau}\,\partial_{\tau} A_{\eta}.
    \label{Eq:P_eta}
\end{equation}
Using the Milne metric together with boost invariance, the classical Yang-Mills equations take the form
\begin{equation}\label{eq:YMEqi}
    \begin{split}
        \partial_{\tau} E^{i}
    	&= \tau\, D_{j} F_{ji}
    	- \frac{ig}{\tau}\,
    	\left[ A_{\eta},\, D_{i} A_{\eta} \right],
        \\
    	\partial_{\tau} E_{\eta}
    	&= \frac{1}{\tau}\, 	D_{i} \left( D_{i} A_{\eta} \right),
    \end{split}
\end{equation}
subject to the Gauss constraint
\begin{equation}
	D_{i} E^{i} + ig\,[A_{\eta}, E_{\eta}] = 0,
	\label{eq:Gauss}
\end{equation}
which remains satisfied throughout the time evolution. 

To ensure exact gauge invariance after discretization in the transverse plane, the above continuum expressions must be rewritten in their lattice Yang-Mills form using link variables and plaquettes.

\subsection{Numerical implementation}			
The evolution of the glasma fields, governed by the CYM equations, is numerically solved within the framework of real-time lattice gauge theory. This approach ensures that the discretized field equations retain the property of gauge invariance. In order to maintain boost invariance at the level of the field configurations, we are not allowed to make $\eta$-dependent gauge transformations. Thus, the field $A_\eta(\tau,\mathbf{x}_\perp)$ transforms as an adjoint-representation scalar, and only the transverse components $A_i$ behave as gauge fields. The transverse plane is taken to be a square of size $L_\perp$, discretized into $N_\perp \times N_\perp$ lattice points with lattice spacing $a_\perp = L_\perp/N_\perp$ and periodic boundary conditions.

We begin with the lattice discretization of the MV model. We decompose each nucleus into $N_s$ independent color sheets along the light-cone. For each sheet $m \in \{1,\dots,N_s\}$, the discretized color charge density $\rho^{a,m}_{L,R}(\mathbf{x}_\perp)$ is sampled from a Gaussian ensemble specified by the one- and two-point correlators
\begin{equation}\label{Eq: MV_Corr_discrete}
    \begin{split}
        \big\langle \rho^{a,m}_{L,R}(\mathbf{x}_\perp) \big\rangle &= 0,
    	\\
    	\big\langle 
    	\rho^{a,m}_{L,R}(\mathbf{x}_\perp)\,
    	\rho^{b,n}_{L,R}(\mathbf{y}_\perp) 
    	\big\rangle
    	&=
    	g^{2}\mu^{2}_{L,R}\;
    	\delta^{ab}\;
    	\frac{\delta^{mn}}{N_s}\;
    	\frac{\delta_{\mathbf{x}_\perp, \mathbf{y}_\perp}}{a_\perp^{\,2}} .
    \end{split}
\end{equation}
At each transverse lattice site $\mathbf{x}_\perp$, for each color component $a$ and each sheet index $m$, the color charge density $\rho^{a,m}_{L,R}(\mathbf{x}_\perp)$ is sampled as an independent Gaussian random variable with zero mean and standard deviation $g\mu_{L,R} /(a_\perp\sqrt{N_s})$, which is consistent with the discretized correlator from Eq.~\eqref{Eq: MV_Corr_discrete}. In this work, we choose $N_s=1$. Usually, the saturation scale $Q_\mathrm{s}$ is defined in terms of the Wilson line correlator. For SU($3$), this discretization leads to an (adjoint-representation) saturation scale $Q_\mathrm{s}\approx 0.6\,g^2\mu$~\cite{Lappi:2007ku}. However, in this first study, our numerical results are for SU($2$). While we expect the physics to be very similar for the physical SU($3$) group, we expect the relation between $Q_\mathrm{s}^2$ and $(g^2\mu)^2$ to be proportional to the quadratic Casimir for the representation.

After generating a random color charge density on the lattice, one has to solve the discretized Poisson equation
\begin{equation}
	\partial_\perp^2 \alpha^{a,m}_{L,R}(\mathbf{x}_\perp) =  -\rho^{a,m}_{L,R}(\mathbf{x}_\perp).
	\label{Eq:Poisson_equation_position_space_discrete}
\end{equation}
This is easiest to do by a Fourier transform, as discussed above.  The natural nearest-neighbor discretization of the Laplacian operator $\partial_\perp^2$ corresponds to replacing the squared momentum in the continuum solution 
in \eqref{Eq: Poisson_Eq_solution_momentum_space_2} by the lattice counterpart
\begin{equation}
	\mathbf{k}_{\perp}^2 \to \tilde{k}_{\perp}^2 = \sum_{i=x,y} \left( \frac{2}{a_\perp} \sin\frac{k_i a_\perp}{2} \right)^2.
	\label{eq:lattice_momentum}
\end{equation} 
Once the fields~$\alpha^{a,m}$ are computed, the Wilson line at transverse position $\mathbf{x}_{\perp}$ is obtained as the ordered product
\begin{equation}
	V^{\dagger}(\mathbf{x}_{\perp})
	=
	\prod_{m=1}^{N_{s}}
	\exp\!\left[-ig\,\alpha_{m}(\mathbf{x}_{\perp})\right],
\end{equation}
where $\alpha^{m} = \alpha^{a,m} T^{a}$. Here, $T^a$ are the $\mathrm{SU}(N_c)$ group generators normalized as $\mathrm{Tr}(T^a T^b) = \delta^{ab}/2$. For $\mathrm{SU}(2)$ or $\mathrm{SU}(3)$  in the fundamental representation, these correspond to the Pauli or Gell-Mann matrices divided by 2.
From the Wilson lines $V_{L,R}(\mathbf{x}_\perp)$, one can then construct link matrices corresponding to the transverse pure gauge single-nucleus fields as
\begin{equation}
	U_{\hati}^{L,R}(\mathbf{x}_\perp)
	=
	V_{L,R}(\mathbf{x}_\perp)\,
	V^{\dagger}_{L,R}(\mathbf{x}_\perp+\hati),
	\label{Eq:GaugeLinks}
\end{equation}
where $\hat{\imath}$ denotes the unit vector in the transverse direction $i\in\{x,y\}$.

The degrees of freedom in our numerical solution of the equations of motion are the transverse components of the gauge field, represented in terms of unitary link matrices, the longitudinal component $A_\eta$, and the electric fields, with the latter all being matrices in the color algebra.

A gauge link $U_{\hati}(\tau, \mathbf{x}_\perp)$ on the lattice represents the Wilson line connecting adjacent lattice sites along the spatial direction $\hati$. In the continuum limit ($a_\perp \to 0$), the link approximates the gauge field as
 \begin{equation}
 	U_{\hati}(\tau, \mathbf{x}_\perp) \approx \exp\left( i g a_\perp A_i \left( \tau, \mathbf{x}_\perp + \frac{a_\perp}{2}\hati \right) \right).
 	\label{Eq:Gauge_Link_Lattice}
 \end{equation} 
Links pointing in the negative transverse direction are obtained from the Hermitian conjugate of the corresponding forward link. Explicitly, for a backward step along $\hati$, one defines
\begin{equation}
	U_{-\hati}(\tau,\mathbf{x}_\perp)
	\equiv
	U^{\dagger}_{\hati}(\tau,\mathbf{x}_\perp - \hati).
	\label{Eq:BackwardLink}
\end{equation}
The continuum initial conditions in Eqs.~\eqref{eq:AiIC} and~\eqref{eq:AetaIC} also need to be implemented in the lattice calculation in terms of the link matrices, which we do in the standard way, as in Refs.~\cite{Krasnitz:1998ns, Muller:2019bwd}.
With the forward and backward links, the elementary plaquette in the $\hati\hatj$-plane is constructed as the ordered product of the gauge links
\begin{equation}\label{Eq:Plaquette}
    U_{\hati\hatj} (\mathbf{x}_{\perp})\equiv U_{\hati} (\mathbf{x}_{\perp})\,
		U_{\hatj} (\mathbf{x}_{\perp}\!+\!\hati)\,U_{\hati}^{\dagger}   (\mathbf{x}_{\perp}  \!+\!\hatj) \,
  		U_{\hatj}^{\dagger}  (\mathbf{x}_{\perp}).
\end{equation}

Using the Yang-Mills action on the lattice, the corresponding canonical momenta are given by
\begin{equation}\label{eq:MomentaLattice}
     \begin{split}
         E_{\eta}(\tau,\mathbf{x}_{\perp})
     	&= \frac{1}{\tau}\,\partial_{\tau} A_{\eta}(\tau,\mathbf{x}_{\perp}),
     	\\[1mm]
     	E^{i}(\tau,\mathbf{x}_{\perp})
     	&= -i\frac{\tau}{\,g a_\perp}
     	\left[
     	\partial_{\tau} U_{\hati}(\tau,\mathbf{x}_{\perp})
     	\right]
     	U_{\hati}^{\dagger}(\tau,\mathbf{x}_{\perp}).
     \end{split}
 \end{equation}
 Varying the lattice Yang-Mills action yields the following discrete equations for the evolution of conjugate momenta:
\begin{align}
	\partial_{\tau} E_{\eta}(\tau,\mathbf{x}_{\perp})
	&=
	\frac{1}{\tau}
	\,D_{i}^{2}A_{\eta}(\tau,\mathbf{x}_{\perp}),
	\label{eq:EOMeta}
	\\[1mm]
	\partial_{\tau} E^{i}(\tau,\mathbf{x}_{\perp})
	&=
	-\sum_{j}
	\frac{\tau}{g a_\perp^{3}}
	\left[
	U_{\hati\hatj}(\tau,\mathbf{x}_{\perp})
	+
	U_{\hati,-\hatj}(\tau,\mathbf{x}_{\perp})
	\right]_{\text{ah}}
\nonumber \\ &
    -i\,\frac{g}{\tau}
	\big[ 
	A^{\text{transp}}_{\eta}(\tau,\mathbf{x}_{\perp}),\,
	D^{F}_{i} A_{\eta}(\tau,\mathbf{x}_{\perp})
	\big].
	\label{eq:EOMi}
\end{align} 
Here, $D^{F}_{i}$ and $D^{B}_{i}$ denote the forward and backward lattice covariant derivatives and $D_{i}^{2}\equiv D^{F}_{i}D^{B}_{i}$. The notation $[\dots]_{\text{ah}}$ denotes $-i$ times the anti-Hermitian traceless part  of a matrix: $[M]_{\text{ah}} \equiv -i [M-M^\dag- \tr(M-M^\dag )/\nc ]/2$. The parallel-transported field is defined as
\begin{equation}
 	A_{\eta}^{\text{transp}}(\tau,\mathbf{x}_{\perp})
 	\equiv
 	U_{\hati}(\tau,\mathbf{x}_{\perp})\,
 	A_{\eta}(\tau,\mathbf{x}_{\perp}+\hati)\,
 	U_{\hati}^{\dagger}(\tau,\mathbf{x}_{\perp}).
 \end{equation}

For the numerical integration, we employ a leapfrog algorithm: the momenta $E_{i}$ and $E_{\eta}$ are evaluated at half-integer time steps $\tau + \Delta\tau/2$, while the fields $U_{\hati}$ and $A_{\eta}$ are updated at integer time steps $\tau$. In this study, we will be measuring fluctuations in the longitudinal electric field $E_\eta$ and in the longitudinal magnetic field $B_\eta$, which we define as follows:
\begin{equation}
	B_{\eta}(\tau_{n},\mathbf{x}_{\perp})
	=
	-F_{xy}(\tau_{n},\mathbf{x}_{\perp})
	\approx
	-\frac{1}{g a_{\perp}^{2}}
	\left[
	U_{\hat{x}\hat{y}}(\tau_n,\mathbf{x}_{\perp})
	\right]_{\text{ah}}.
	\label{Eq: B_eta_Physical}
\end{equation}

With all these ingredients, the glasma fields are evolved forward in proper time. For this, we use the numerical real-time lattice solver originally developed in Refs.~\cite{Muller:2019bwd, Ipp:2020mjc, Ipp:2020nfu} and later also used in Refs.~\cite{Avramescu:2023qvv, Avramescu:2024poa, Avramescu:2024xts}. As we take $N_s=1$ and do not use a gluonic mass regulator, our glasma fields can depend on two dimensionless combinations of the saturation scale $g^2\mu$, the lattice spacing $a_\perp$, and the lattice volume $L_\perp^2$, and we must check the dependence of our results on these ratios. Furthermore, using the aforementioned numerical solver, we perturb the electric and magnetic fields and measure the growth rate of the perturbations using various scenarios, with our simulation routines publicly available\footnote{The simulation code for the glasma fields and extraction of the Lyapunov exponents is available at~\href{https://github.com/pooja4593/Glasma/tree/Glasma\_Lyapunov}{\texttt{github.com/pooja4593/Glasma/tree/Glasma\_Lyapunov}}.}.


\section{Perturbation to the fields}
\label{Section: Formalism_Perturbation}
We measure how small perturbations grow over time in Yang-Mills field simulations. If the growth is exponential, its rate, the Lyapunov exponent,  characterizes the chaotic nature of the system. To do this in practice, we initialize two nearly identical field configurations at $\tau=0$, and monitor the growth of the difference between them as a function of proper time $\tau$. By starting from a fluctuation that is small, and following the time evolution long enough, we become sensitive to only the fastest growing mode of the fluctuations, the leading Lyapunov exponent. We will extract it by fitting the time dependence of the growth at late times.

The nonzero gauge field components at $\tau=0$ are the longitudinal electric field $E_\eta$ and the longitudinal magnetic field $B_\eta$. These correspond to different polarization states of gluons in the glasma, and in a free theory, would not interact with each other. If we observe exponential growth in the perturbation, it is not a priori obvious whether the rate would be the same for both components. Thus, by perturbing either one of them separately, and then measuring both, we can infer whether the fastest growing exponential couples to both or just one of these polarization states, providing information on how different field components contribute to overall chaotic behavior. We need to make sure that the initial perturbation amplitude is kept sufficiently small so that the system remains in the linear regime.

We want to have access to the way unstable fluctuations couple to different momentum modes of the gauge field. For that purpose, it is useful to think of the initialization of the perturbations in momentum space. We will first initialize the perturbations with random white noise, which perturbs all momentum modes present on the lattice. We will then modify this with a spectral modulation that restricts the noise to only some momentum modes. We now discuss these two options in more detail.

\subsection{White noise perturbation}
\label{Subsection: White_Noise}   

A white noise perturbation to the longitudinal electric field is introduced in the following manner.
The perturbed field $E_\eta^\prime$  is constructed from the original longitudinal field $ E_\eta$ as follows
\begin{equation}
		E_\eta^{\prime a}(\tau=0, \mathbf{x}_{\perp}) = E_\eta^a (\tau=0, \mathbf{x}_{\perp}) + \delta E_\eta^a (\tau=0, \mathbf{x}_{\perp}),
\end{equation}
with $\delta E_\eta^a(\mathbf{x}_{\perp}) = 2 \chi^a(\mathbf{x}_{\perp}) / a_{\perp}^{2}$. The term $\chi^a(\mathbf{x}_{\perp})$ is a Gaussian random variable sampled independently for each transverse lattice site and color index. It follows a normal distribution with zero mean and a variance $\alpha^2$ such that
\begin{equation}\label{Eq:chi_Moments}
    \begin{split}
        \langle  \chi^a(\mathbf{x}_{\perp}) \rangle &= 0, \\
    	\langle \chi^a(\mathbf{x}_{\perp}) ~\chi^b(\mathbf{y}_{\perp}) \rangle &= \alpha^2 \delta^{ab}  ~ \!  \! \delta_{\mathbf{x}_{\perp} , \mathbf{y}_{\perp}}.
    \end{split}
\end{equation}
Here, $\alpha$ is a small, real-valued parameter that controls the overall amplitude or strength of the perturbation, $\delta^{ab}$ is the Kronecker delta over the color indices, ensuring that the noise is uncorrelated across different generators of the gauge group and $\delta_{\mathbf{x}_{\perp}, \mathbf{y}_{\perp}}$ ensures that the noise is uncorrelated between different points in space. To study the chaotic behavior via Lyapunov exponents, $\alpha$ is chosen to be sufficiently small (typically $\alpha  \ll 1$) to ensure that the evolution of the perturbation, measured via $\langle \tr {(\delta E_\eta)^2}\rangle$ remains within the linear regime, i.e., scales as $\alpha^2$ at all times.

To study the growth of fluctuations in the magnetic field sector, we need to perturb the fundamental gauge fields (or gauge links) directly, since the magnetic field is not, in itself, a dynamical variable in the equations of motion. To this end, we define the perturbed transverse gauge fields $A_i^{\prime a}$ at the initial proper time as
\begin{equation}
	A_i^{\prime a}(\tau=0, \mathbf{x}_{\perp}) = A_i^a(\tau=0, \mathbf{x}_{\perp}) + \delta A_i^a (\tau=0, \mathbf{x}_{\perp}),
\end{equation}
where $\delta A_i^a$ represents the stochastic perturbation to the gauge field components in the transverse plane.
On the lattice, this translates to a perturbation of the corresponding gauge links $U_{i}^{\prime}$ as
\begin{multline}    
	U_{\hati}^{\prime}(\tau=0, \mathbf{x}_{\perp}) = U_{\hati} (\tau=0, \mathbf{x}_{\perp})
\\
  \times \exp \left[ i g a_\perp ~ \! \! \delta A_i^a
	\left( \tau=0, \mathbf{x}_{\perp}  \right) T^a \right] .	
\end{multline}
The phase fluctuation $\delta A_i^a$ is introduced as a stochastic noise term, defined by $\delta A_i^a (\mathbf{x}_{\perp}) = 2 \zeta_i^a (\mathbf{x}_{\perp}) / a_{\perp}$. 
Similar to $\chi^a$, the term $\zeta^a_i$ is a Gaussian random variable sampled independently for each transverse direction $i\in\{x,y\}$, lattice site, and color index, satisfying
\begin{equation}\label{Eq:zeta_Moments}
    \begin{split}
        \langle  \zeta^a_i(\mathbf{x}_{\perp}) \rangle &= 0, \\
    	\langle \zeta^a_i(\mathbf{x}_{\perp}) ~\zeta^b_j(\mathbf{y}_{\perp}) \rangle &= \beta^2 \delta^{ab} \delta_{ij} ~ \! \! \delta_{\mathbf{x}_{\perp} , \mathbf{y}_{\perp}}.
    \end{split}
\end{equation}
Here, $\beta$ is a small parameter controlling the strength of the longitudinal magnetic field perturbation, analogous to $\alpha$ in the electric field case. Next, the perturbed links are then used to construct the perturbed transverse plaquettes $U_{\hati\hatj}^{\prime}$. These plaquettes are subsequently substituted into the definition of the physical longitudinal magnetic field, namely, Eq.~\eqref{Eq: B_eta_Physical}, to get $B_\eta^{\prime}$,  with the difference $\delta B_\eta = B_\eta^{\prime} - B_\eta$ representing the initial magnetic field perturbation.
This procedure effectively seeds a perturbation in $B_\eta$ while maintaining the unitarity of the gauge links. It is important to note that because the noise $\zeta^a_i$ is sampled independently for each link (i.e., each lattice site and each direction), this perturbation does not correspond to a gauge transformation. Instead, it represents physical fluctuations in the field configuration, which is the required condition for studying chaotic dynamics and extracting Lyapunov exponents.

As with the electric field case, we ensure the perturbation remains in the linear regime by selecting $\beta \ll 1$. Typical values used in our simulations are $\alpha, \beta \sim 10^{-5} - 10^{-3}$, ensuring that non-linear effects from the perturbation itself are negligible compared to the intrinsic chaotic dynamics of the system.
It should be noted that the appropriate power of the lattice spacing has been scaled out so that $\chi^a, \zeta^a_i$ are dimensionless variables.

\subsection{Spectral modulation and momentum filtering}
\label{Subsection: Spectral_Modulation}

Next, to understand how the chaotic dynamics depend on the spatial scale of the initial perturbations, we introduce momentum-space filtering that selectively perturbs specific Fourier modes of the field configurations. Instead of purely local white noise, the perturbations are modulated by a noise kernel $\mathcal{K}_{\text{f}}(k_{\perp})$ that depends on the magnitude of the transverse momentum $k_{\perp} = |\mathbf{k}_{\perp}|$. This allows us to study whether the chaotic behavior exhibits scale dependence and to distinguish between ultraviolet and infrared sensitivity in the Yang-Mills system.

The filtered perturbation $\psi_{\text{f}}(\mathbf{x}_{\perp})$ is obtained by convolving the initial Gaussian white noise $\psi(\mathbf{x}_{\perp}) \in\{\chi^a, \zeta^a_i\} $ with a kernel function. To do this, firstly, the spatial noise field is transformed to momentum space as
\begin{equation}
	\widetilde{\psi} (\mathbf{k}_{\perp}) = \mathcal{F}  \left[  \psi(\mathbf{x}_{\perp}) \right]  = \sum_{\mathbf{x}_{\perp}} e^{-i\mathbf{k}_{\perp}\cdot\mathbf{x}_{\perp}} \psi(\mathbf{x}_{\perp}),
\end{equation}
where $\mathcal{F}$ denotes the discrete Fourier transform. The Fourier components are multiplied by a filter function: 
\begin{equation}
	\widetilde{\psi}_{\text{f}}(\mathbf{k}_{\perp})  = \mathcal{K}_{\text{f}}({k}_{\perp}) ~ \! \widetilde{\psi} (\mathbf{k}_{\perp}).
\end{equation}
The filtered noise is transformed back to position space as
\begin{equation}
	\psi_{{\text{f}}}(\mathbf{x}_{\perp}) = \mathcal{F}^{-1} \left[ \widetilde{\psi}_{\text{f}}(\mathbf{k}_{\perp})  \right],
\end{equation}
where $\mathcal{F}^{-1}$ denotes the inverse discrete Fourier transform. This filtered noise is then used in place of the original white noise in the perturbation procedures described previously.

\textbf{Noise kernel profiles:} To control the localization of the perturbation in momentum space and investigate the scale dependence of the chaotic dynamics, we implement three distinct types of kernels $\mathcal{K}_\mathrm{f}(k_\perp)$. These kernels modulate the white noise in Fourier space before it is transformed back to the position-space lattice.

\begin{enumerate}
\item \textbf{Gaussian filter}: This filter suppresses high-momentum (short-distance) modes, creating perturbations that are correlated over a characteristic length scale $\xi = 1/\kappa_\mathrm{g}$. The filter function is 
\begin{equation}
	\mathcal{K}_{\text{gauss}}(k_\perp) = \exp\left( -\frac{k_\perp^2}{\kappa_\mathrm{g}^2} \right).
\label{Eq:Gaussian_Filter}
\end{equation}	
where $\kappa_\mathrm{g}$ is the filter mass parameter that controls the width of the Gaussian. Smaller values of $\kappa_\mathrm{g}$ correspond to aggressive filters that suppress more high-momentum modes.

\item \textbf{Power-law filter}: This filter creates perturbations with a power-law momentum dependence.  Compared to the Gaussian case, this kernel allows for a heavier tail in the UV region, thus retaining more dependence from high-momentum fluctuations. The filter function is
\begin{equation}
	\mathcal{K}_{\text{power}}(k_\perp) = \frac{\kappa_\mathrm{p}^2}{k_{\perp}^2 + \kappa_\mathrm{p}^2}
	\label{eq:power_filter}.
\end{equation}
The parameter $\kappa_\mathrm{p}$ acts as a cutoff scale above which momentum modes start to get suppressed.

\item \textbf{Shell (band-pass) filter}: To isolate the contribution of specific momentum modes to chaotic growth, we use a step-function filter that selects modes within a narrow band of momenta:
\begin{equation}
	\mathcal{K}_{\text{shell}}({k}_{\perp}) = 
	\begin{cases}
		1, & \text{if } \kappa_\mathrm{s} - \Delta k/2 \leq k_{\perp} \leq \kappa_\mathrm{s} + \Delta k/2 \\
		0, & \text{otherwise}
	\end{cases}
	\label{Eq:Shell_Filter}
\end{equation}
Here, $\kappa_\mathrm{s}$ is the central momentum of the shell, and we take its width to be  $\Delta k = 2(2\pi/L_\perp)$ so that enough discrete lattice momentum modes are included in the shell. 
\end{enumerate}

As a side note, let us remark that after applying a Gaussian or power-law filter, the ensemble-averaged squared fluctuation of the longitudinal electric field at the initial time is
\begin{equation}
\langle \tr {(\delta E_\eta (\tau=0) )^2}\rangle   \propto   \frac{\kappa^2}{a_{\perp}^2},
	\label{Eq:Eeta_growth_scaling}
\end{equation}
with $\kappa \in \{ \kappa_\mathrm{g}, \kappa_\mathrm{p} \}$. We will need this to properly analyze the dependence of our results on the lattice discretization. The dimensionless quantity $a_{\perp}^4 \langle \tr (\delta E_\eta)^2 \rangle \propto \kappa^2 L_\perp^2/N_\perp^2$ will need to be scaled with $N_\perp^2$ to properly compare simulations with different lattice spacings $a_{\perp}$ and a fixed volume $L_\perp^2$.


\begin{figure}[tb!]
\hfill \includegraphics[height=7cm]{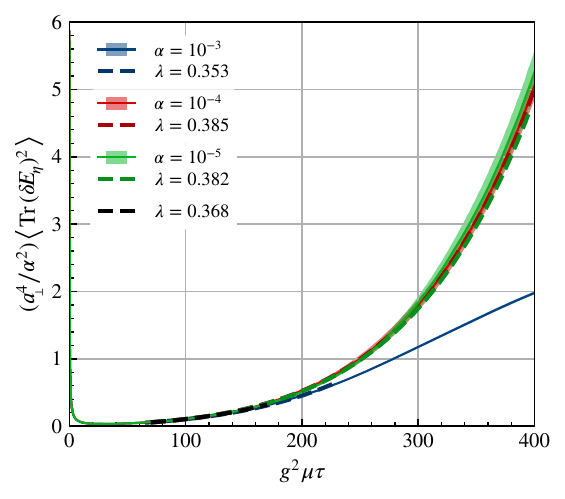}
\hfill \includegraphics[height=7cm]{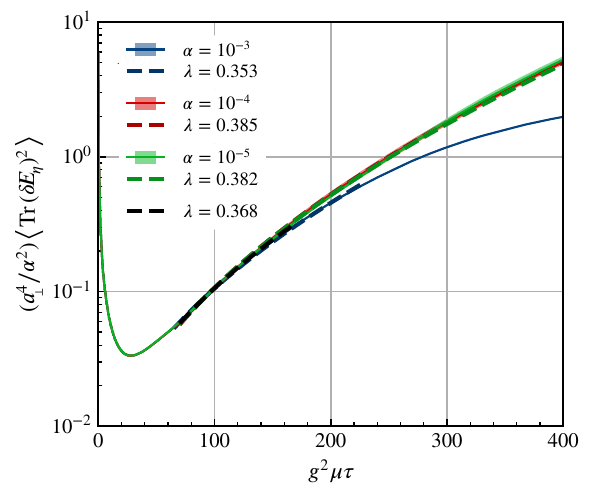}
\caption{
 Rescaled time $g^2\mu\tau$ dependence of the longitudinal electric field perturbation $\langle \mathrm{Tr}(\delta E_\eta)^2\rangle$ with a white noise initial condition as given in Eq.~\eqref{Eq:chi_Moments}, shown both on a linear and logarithmic scale. The solid lines and associated error bands represent the average and statistical errors of the data, with different colors for different values of $\alpha$. The corresponding colored dashed lines are the separate fits to these datasets, and the black dashed line is a fit to the combination of the datasets.
 The extracted Lyapunov exponent $\lambda$ for each fit is provided in the legend.
The fit ranges given by our procedure and the resulting fits are given in Table~\ref{Table: NoFilter} in Appendix~\ref{Appendix:Fit_Results}. 
}	
\label{Fig:Figure_NoNoise}
\end{figure}

\section{Extracting the Lyapunov exponent}
\label{Section: Lyapunov_Extraction}
In this section, we begin our analysis by investigating the growth of electric field perturbations using the spatially uncorrelated white noise described in Sec.~\ref{Subsection: White_Noise}, which serves as a baseline case without momentum-space filtering. We use this result to develop our fitting procedure to extract the leading Lyapunov exponent. Then, we will present our main results in Sec.~\ref{Section: Results}, exploring how the chaotic response is modified by the introduction of different momentum-space filters, namely the Gaussian, power-law, and shell filters of Sec.~\ref{Subsection: Spectral_Modulation}, comparing electric-field and magnetic-field fluctuations under the same filtering conditions and investigating the dependence of our results on the lattice parameters  $g^2\mu a_\perp$ and  $g^2\mu L_\perp$. 

In this study, we have only used the SU($2$) color group in order to be able to use larger lattices and accumulate more statistics. We do not expect the qualitative features of the physics to be different from SU($3$).
Unless stated otherwise, we use a $N_\perp=256$ lattice with $g^2\mu a_\perp = 0.469$ and  $g^2\mu L_\perp = 120$. The numerical calculation only depends on these dimensionless combinations rather than $g^2\mu$, $a_\perp$, and  $L_\perp$ separately. To turn them into values in physical units, one could, for example, take $L_\perp \approx 23.68$~fm, which would correspond to a lattice spacing $a_\perp \approx 0.093$~fm and $g^2\mu = 1$~GeV, noting that we work here in natural units $\hbar = c = 1$. 
Throughout this paper, the ensemble statistics are accumulated from $N_\mathrm{events} = 5000$ independent configurations. This is much larger than the number of configurations typically needed for studies of the glasma fields.  We have found that this larger statistic is required to provide robust statistical precision for the extracted Lyapunov exponents. Unless stated otherwise, we show results for the electric field fluctuations in response to initial electric field perturbations.

The Lyapunov exponent is extracted from the exponential growth of field differences
\begin{equation}
    \langle \tr {(\delta F_\eta(\tau) )^2}\rangle \propto e^{\lambda  \sqrt{g^2\mu \tau}},
\end{equation}
where $F_\eta = E_\eta, B_\eta$ denotes the longitudinal field component being perturbed. These Lyapunov exponents are extracted at a sufficiently late stage when the system is dominated by a single exponentially growing mode.

To ensure that our simulations probe the intrinsic chaoticity of the system rather than non-linear artifacts, we select perturbation strengths $\alpha, \beta \ll 1$. This choice maintains the system within the linear-response regime, which we verify through two primary criteria. 
First, we confirm that the ensemble-averaged fluctuations, $\langle \mathrm{Tr}(\delta E_\eta)^2 \rangle$ and $\langle \mathrm{Tr}(\delta B_\eta)^2\rangle$, exhibit a clear exponential growth over a sustained period of proper time. Second, we perform a scaling test to ensure that the measured fluctuations are proportional to the square of the initial perturbation amplitude; specifically, doubling the value of $\alpha$ or $\beta$ results in a fourfold increase in the magnitude of the fluctuations, as expected in the linear regime.

\subsection{Time dependence of the fluctuations}
\label{Subsection: No_Noise}
The growth of longitudinal electric field fluctuations in response to electric field perturbations is presented in Fig.~\ref{Fig:Figure_NoNoise}. To verify the linearity of the response, we plot the ensemble-averaged fluctuations $a^4_{\perp}\langle\mathrm{Tr}(\delta E_\eta)^2\rangle$ normalized by the square of the perturbation amplitude $\alpha^2$ as a function of the dimensionless proper time $g^2\mu\tau$. We show the data on both linear and logarithmic scales to highlight the exponential nature of the growth. We consider three distinct perturbation strengths: $\alpha \in\{ 10^{-3}, 10^{-4}, 10^{-5}\}$, spanning three orders of magnitude. Each dataset comprises a solid line showing the mean evolution and a lightly shaded band representing the statistical uncertainty from ensemble averaging over multiple independent configurations. The figure also shows best-fit curves, obtained through a procedure that we will describe shortly, as thicker dashed lines. The same style will also be used in other figures in this paper. 

The first observation is that after a small initial transient time ($g^2\mu\tau \sim 40$), the fluctuations enter a clear exponential growth regime. This growth can be very accurately characterized by the functional form 
\begin{equation}
\frac{a^4_{\perp}}{\alpha^2} \left\langle	\mathrm{Tr}(\delta E_\eta)^2\right\rangle = a \exp\left(\lambda\sqrt{g^2\mu\tau}\right),
	\label{eq:growth_form}
\end{equation}
where the dimensionless number $\lambda$ is what we will here refer to as the  Lyapunov exponent, in this case for electric field perturbations. A crucial observation is that the fit form works well as long as all three curves collapse when normalized by $\alpha^2$. At late times, the data for the largest value, $\alpha=10^{-3}$, start to deviate from linear behavior.  This scaling behavior confirms that by choosing small enough values for the normalization $\alpha$, we can stay in a regime where the response of the system is linear in $\alpha$, ensuring that the extracted Lyapunov exponents are intrinsic properties of the glasma dynamics rather than artifacts of the perturbation magnitude. Let us now discuss the fitting procedure used to obtain the growth rates from the time dependence of the perturbations.

\subsection{Fitting procedure}
\label{Subsection: Fitting_Procedure}
To extract the Lyapunov exponents from the time-evolving field fluctuations, we perform a robust multi-range regression analysis. The growth of the field fluctuations, denoted by $y(t)$, is seen to follow the functional form 
\begin{equation}
	y(t) = a \exp\left( \lambda \sqrt{t} \right),
	\label{Eq: FittingForm}
\end{equation}
where $a$ is a perturbation amplitude factor, $\lambda$ represents the Lyapunov exponent, and $t = g^2 \mu \tau$ is the dimensionless proper time.

We want to develop a systematic scanning-window technique to identify the optimal fitting interval $[t_{\min}, t_{\max}]$ that starts after the initial transients have subsided and ends before the perturbation amplitude enters the non-linear regime. Our method aims to balance two competing requirements. A short fitting interval will provide a good $\chi^2$, but it will not constrain the exponent very effectively. An overly long fitting interval, on the other hand, will extend beyond the validity of the fitting form.  To eliminate subjectivity in choosing the fitting window, we perform an exhaustive scan over a grid of minimum and maximum times: $t_{\min} \in [5,100]$ and $t_{\max} \in [30,400]$, with a step size of $\Delta t = 5$.
For each given interval, we perform a weighted non-linear least-squares fit of the form given in Eq.~\eqref{Eq: FittingForm} to the numerical data, accounting for the statistical errors. The quality of each fit is assessed via the reduced chi-squared statistic $\chi^2_{\mathrm{red}} = \chi^2/(N - 2)$, where $N$ is the number of data points within the fitting interval and $2$ corresponds to the number of free parameters ($a$ and $\lambda$). 
The selection of the optimal window is governed by a hierarchical process. We first identify a pool of statistically consistent fits, defined by the condition $\chi^2_{\mathrm{red}} < 1$.
From this subset of acceptable fits,  the best fit is selected as the one that minimizes the standard error of the exponent, $\delta \lambda$. This criterion ensures that our estimate of $\lambda$ has the smallest statistical uncertainty among all statistically acceptable fits, ensuring that the extracted $\lambda$ is precisely determined over a stable dynamical range. To guarantee that the extracted exponent is not an artifact of a narrow temporal window, we verify that the selected interval $[t_{\min}, t_{\max}]$ spans a sufficient dynamical range for the exponential growth to be reliably established. In the rare case where no fit satisfies $\chi^2_{\mathrm{red}} < 1$, we employ a fallback strategy: we select the fit with the smallest $\chi^2_{\mathrm{red}}$ value. This ensures that we always obtain a result, albeit with an appropriate note of caution regarding the statistical quality of the fit.

We emphasize that the statistical errors from the independent configurations are used in our fitting procedure as described above. However, we have enough events that the statistical error $\delta \lambda$ from individual fits is practically always smaller than the differences between fits to different datasets.   
As a consequence, when we quote uncertainties for the values of $\lambda$  in this paper, they result from a rough comparison between extractions using different initializations of the perturbation and should be understood as an estimated systematic uncertainty rather than a statistical one.

Returning to Fig.~\ref{Fig:Figure_NoNoise}, it also shows the results of our systematic fitting procedure applied to the growth of electric field perturbations, with fit ranges and parameters summarized also in Table~\ref{Table: NoFilter} in Appendix~\ref{Appendix:Fit_Results}. To obtain the most precise estimate of the Lyapunov exponent, we analyze both individual datasets for each perturbation strength $\alpha$ and a combined dataset that aggregates statistics across all three perturbation strengths, and fit them separately using the method described above.

Applying the aforementioned fitting procedure to each dataset yields consistent values of the Lyapunov exponent $\lambda$ across all perturbation strengths. We find that a value $\lambda \approx 0.37$ corresponds to the fit results for the  separate  datasets $\alpha \in\{ 10^{-3}, 10^{-4}, 10^{-5}\}$ and the combined dataset. Our fit procedure manages to automatically exclude the late time region for $\alpha=10^{-3}$, where the nonlinear behavior in the fluctuations causes the dataset to deviate from the fit form. 

The main result of Fig.~\ref{Fig:Figure_NoNoise} is that the growth rate extracted with our fitting procedure is found to be independent of the normalization $\alpha$, as long as we limit ourselves to small enough values of $\alpha$. The linearity in $\alpha$ provides strong evidence that we are probing the intrinsic chaotic dynamics of the system rather than non-linear effects dependent on the perturbation strength. Furthermore, while the primary focus here is on the longitudinal electric field sector, we verified that the longitudinal magnetic field fluctuations for the same lattice exhibit a similar growth pattern with a consistent exponent. Therefore, we use the same fitting procedure for the magnetic field perturbations as well. 

\section{Results}
\label{Section: Results}
Next, to understand whether the chaotic growth of classical Yang-Mills fields exhibits any intrinsic scale dependence, we examine the evolution of perturbations filtered in momentum space that selectively excite specific momentum modes. By restricting the initial longitudinal electric field perturbation to specific momentum modes, we can determine if the observed instabilities are predominantly driven by infrared or ultraviolet modes.
In this work, we employ three distinct filter models: Gaussian, power-law, and shell filters, as described in Sec.~\ref{Subsection: Spectral_Modulation}, to probe the spectral response of the glasma.

\subsection{Gaussian filtered perturbations}
\label{Subsection: Results_Gaussian}

\begin{figure}[t]
\centering
\includegraphics[width=0.47\textwidth]{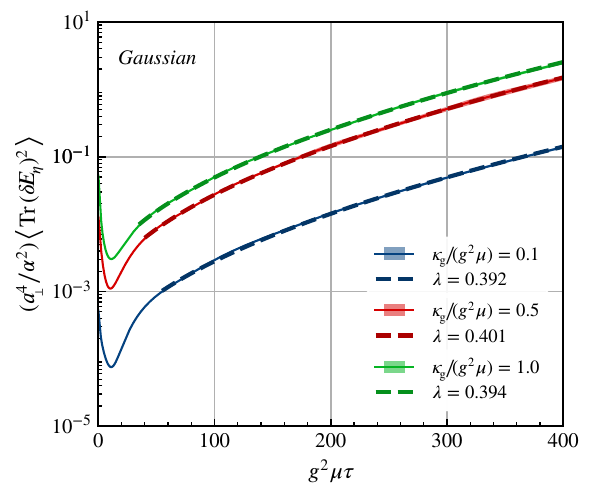}
\caption{
The electric field perturbation $\langle \mathrm{Tr}(\delta E_\eta)^2\rangle$ as a function of $g^2\mu\tau$ with Gaussian filtered noise according to Eq.~\eqref{Eq:Gaussian_Filter}. 
The evolution is shown for different filter scales $\kappa_\mathrm{g}/(g^2\mu)$ shown in a different color, with the same perturbation amplitude $\alpha = 10^{-4}$. The resulting fit parameters are given in Table \ref{Table: Gaussian_Filter} in Appendix~\ref{Appendix:Fit_Results}.
}
\label{Figure: Gaussian_Filter}
\end{figure}

The Gaussian filter is used to study the impact of a soft momentum cutoff on the growth rate. 
Fig.~\ref{Figure: Gaussian_Filter} presents the growth of electric field perturbations $\langle \mathrm{Tr}(\delta E_\eta)^2\rangle$ as a function of proper time $g^2\mu\tau$ on a semi-logarithmic scale.
Also shown are the fits obtained using the procedure described in Sec.~\ref{Subsection: Fitting_Procedure}, with the fit parameters given in  Table~\ref{Table: Gaussian_Filter} in Appendix~\ref{Appendix:Fit_Results}.
Here, we apply Gaussian-filtered $E_\eta$ perturbations, given by  Eq.~\eqref{Eq:Gaussian_Filter} with three representative filter scales: $\kappa_\mathrm{g}/(g^2\mu) \in\{ 0.1, 0.5, 1.0\}$,  where $\kappa_\mathrm{g}$ serves as a characteristic momentum cutoff. Smaller values of $\kappa_\mathrm{g}$ suppress high-momentum modes more strongly and therefore generate perturbations dominated by infrared fluctuations, while a larger $\kappa_\mathrm{g}$ allows more of the ultraviolet modes to pass.
 
As discussed below, the momentum space filter decreases the overall amplitude of the fluctuations compared to the white noise case. While for the largest amplitude $\alpha=10^{-3}$ for the white noise perturbation in Fig.~\ref{Fig:Figure_NoNoise}  we saw signs of non-linear behavior at the latest times, the Gaussian filtering makes the linear behavior more visible, which we have verified with separate simulations using different values of $\alpha$.  Having established the linearity in $\alpha$ and $\beta$, we show in Fig.~\ref{Figure: Gaussian_Filter} and subsequent plots the time dependence of the perturbation only for  $\alpha =\beta = 10^{-4}$ for visual clarity. However, fit results from different values of  $\alpha$ are listed in the tables in Appendix~\ref{Appendix:Fit_Results}. A similar plot style and the same lattice configuration will be used for the different filters presented later in this section.

We observe that all three datasets exhibit clean exponential growth after the transient phase. The statistical errors are quite small, and the fit form works very well.  Applying our fitting methodology, we extract Lyapunov exponents $\lambda \approx 0.40$, irrespective of the value of $\kappa_\mathrm{g}/(g^2\mu)$ used. These values are consistent with the unfiltered result within the stated uncertainties. 

While the exponential growth rates are nearly identical, the amplitudes of the perturbations differ across the three cases. This variation arises from the filtering process itself. For smaller values of $\kappa_\mathrm{g}$, the filter suppresses a larger portion of the momentum spectrum, resulting in fewer modes contributing to the initial fluctuation. Hence, we get a smaller initial intercept on the y-axis. On the other hand,  larger $\kappa_\mathrm{g}$ values allow higher-momentum modes to contribute, resulting in a larger initial amplitude as more independent Fourier modes participate in the perturbation.

The observed scale independence of $\lambda$ across filter scales spanning an order of magnitude carries significant physical implications. It suggests that chaos in the glasma is not driven by specific instabilities localized at particular momentum scales, but rather represents a collective, scale-invariant property of the  Yang-Mills dynamics. The non-linear interactions in the classical gauge field theory appear to couple all spatial scales efficiently.

\begin{figure}[t]
\centering
\includegraphics[width=0.47\textwidth]{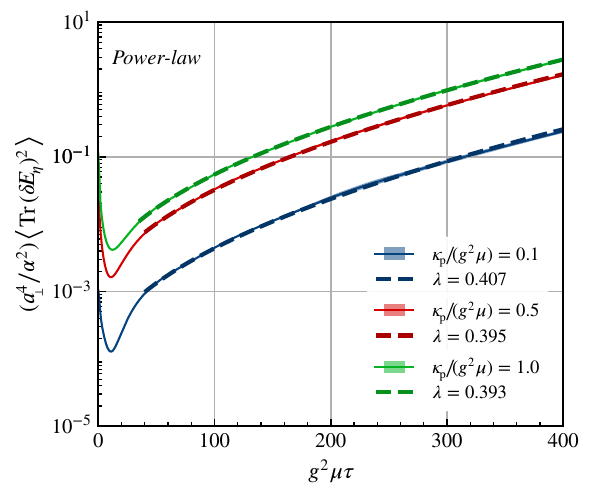}
\caption{
The electric field perturbation $\langle \mathrm{Tr}(\delta E_\eta)^2\rangle$ as a function of $g^2\mu\tau$ with power-law filtered noise given in Eq.~\eqref{eq:power_filter}. 
The evolution is shown for different filter scales $\kappa_\mathrm{p}/(g^2\mu)$  shown in a different color, with the same perturbation amplitude $\alpha = 10^{-4}$. The resulting fit parameters are given in Table \ref{Table: Powerlaw_Filter} in Appendix~\ref{Appendix:Fit_Results}.
}
\label{Fig: Powerlaw_Filter}
\end{figure}


 \begin{figure*}[ht!]
	\centering
\begin{minipage}{0.47\linewidth}
	\centering 	\includegraphics[width=\linewidth]{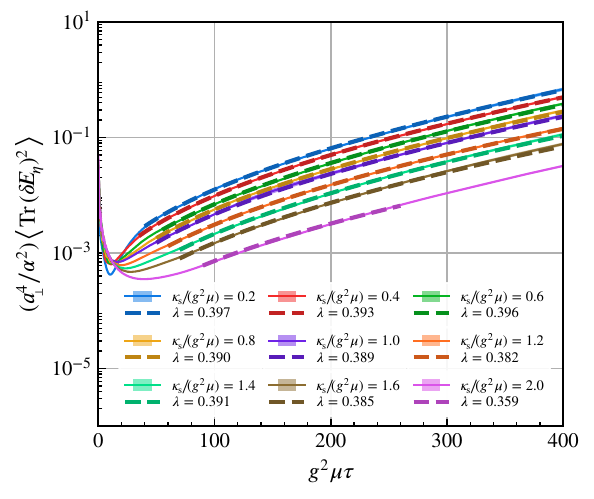}
\end{minipage}
\hfill
\begin{minipage}{0.47\linewidth}
	\centering
	\includegraphics[width=\linewidth]{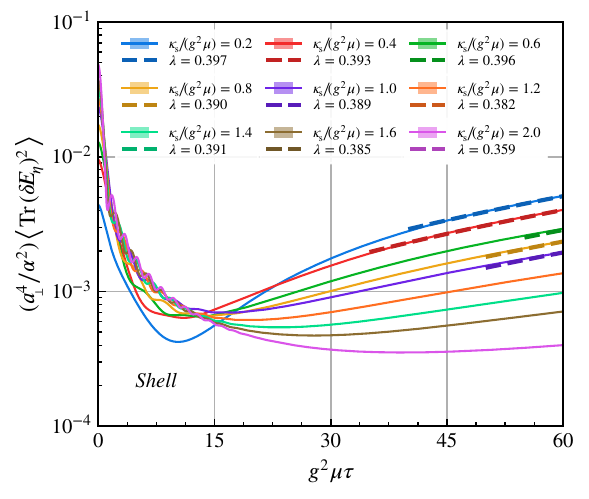}
\end{minipage}	
\caption{The electric field perturbation $\langle \mathrm{Tr}(\delta E_\eta)^2\rangle$ as a function of time $g^2\mu\tau$ for different momentum shell initializations for $\alpha=10^{-4}$, together with the corresponding fits. The plot on the left shows the full simulation interval, and the one on the right zooms in to the early time behavior. 
}
 \label{Fig:shelltimedep}
 \end{figure*}

\subsection{Power-law filtered perturbations}
\label{Subsection: Results_Powerlaw}

To further test the robustness of scale-independent chaotic growth, we examine perturbations filtered with a power-law kernel given by Eq.~\eqref{eq:power_filter}. Unlike the Gaussian filter, which exponentially suppresses high-momentum modes, the power-law filter has a slower, algebraic fall-off, thereby preserving more of the ultraviolet tail while still introducing a characteristic scale $\kappa_\mathrm{p}$. This enables us to probe whether the chaotic dynamics are sensitive to the detailed shape of the momentum-space perturbation.

Figure~\ref{Fig: Powerlaw_Filter} displays the growth of electric field perturbations for power-law filtered initial electric perturbations, evaluated at the same three filter scales as in the Gaussian case: $\kappa_\mathrm{p}/(g^{2}\mu)\in\{ 0.1, 0.5, 1.0\}$ on a semi-logarithmic scale for $\alpha = 10^{-4}$. 
The fit parameters are given in  Table \ref{Table: Powerlaw_Filter} in Appendix~\ref{Appendix:Fit_Results}. As shown in the figure, the behavior of the power-law-filtered perturbations closely mirrors that of the Gaussian case. Despite the different spectral shape of the initial seed, the system quickly enters an exponential growth regime. Our fitting procedure yields Lyapunov exponents that are again centered around $\lambda \approx 0.39$ for all three scales. The persistence of this specific value is notable. While the power-law filter allows more high-momentum modes than the Gaussian filter for the same value of $\kappa_\mathrm{p}$, the growth rate remains invariant.

Similar to the Gaussian filter, variations in the filter scale $\kappa_\mathrm{p}$ primarily affect the initial amplitude of the perturbation. For smaller values of $\kappa_\mathrm{p}$, the higher suppression of higher momentum modes reduces the number of contributing Fourier modes, resulting in a smaller initial perturbation norm.
Conversely, larger values of $\kappa_\mathrm{p}$ allow more ultraviolet modes to participate, leading to a larger initial amplitude. Importantly, these differences do not affect the slope of the exponential growth on the logarithmic scale, and the three datasets remain nearly parallel throughout the growth regime.
The agreement of the Lyapunov exponents across different power-law filter scales further strengthens the conclusion from the Gaussian filter, namely, that chaos in the glasma is a collective phenomenon governed by the non-linear Yang-Mills dynamics rather than by instabilities tied to specific momentum scales.

\subsection{Momentum shell filtered perturbations}
\label{Subsection: Results_Shell}

The final test of scale independence uses a momentum shell filter. 
Unlike the previous filters that only provide smooth ultraviolet cutoffs, the shell filter isolates the initial perturbation to a narrow range of momentum modes centered around a specific value $\kappa_\mathrm{s}$, defined by Eq.~\eqref{Eq:Shell_Filter}. This allows us to probe whether specific momentum shells act as primary drivers for the observed instabilities. A collection of fit parameters for the momentum shell perturbations is given in Table \ref{Table: Shell_Filter} in Appendix~\ref{Appendix:Fit_Results}.

Figure~\ref{Fig:shelltimedep} displays the growth of electric field fluctuations for different momentum shells perturbations on a semi-logarithmic scale, with the left panel displaying the full time evolution and the right panel focusing on the early-time behavior for clarity. We explore nine distinct momentum scales spanning from infrared to the ultraviolet regimes: $\kappa_\mathrm{s} /(g^2\mu) \in \{0.2, 0.4, 0.6, 0.8, 1.0, 1.2, 1.4,  1.6, 2.0\}$, with the perturbation amplitude fixed at $\alpha = 10^{-4}$. 
 
We observe a clear vertical hierarchy in the growth curves. At initial times, for a fixed perturbation amplitude $\alpha$, higher momentum shells (larger $\kappa_\mathrm{s}$) exhibit larger initial intercepts on the logarithmic scale compared to lower-$\kappa_\mathrm{s}$ shells. This is due to the larger phase-space area $A \sim 2\pi \kappa_\mathrm{s}\Delta k$ available at higher momenta, which allows more Fourier modes to contribute to the initial energy density of the perturbation. At early times, the fluctuation modes exhibit an oscillatory behavior at a frequency increasing with their momentum. We attribute this to a free-field-like behavior, which is explicit in a perturbative linearized solution to the equations of motion~\cite{Guerrero-Rodriguez:2021ask}.
 Nevertheless, after an initial dilution phase driven by longitudinal expansion, the oscillations die away, and all perturbations exhibit exponential growth. 
Contrary to the early-time ordering, the late-time amplitude of the unstable mode is smaller for perturbations initialized at higher momentum. This indicates that while the high momentum modes still do couple to the dominant unstable mode, the overlap is smaller.

We see that the unstable mode has a nontrivial momentum-space structure, rather than being associated with a single momentum scale.
A more detailed characterization would potentially require a Fourier-space analysis of the perturbation.
Since such an analysis is not gauge-invariant, we defer discussing it to future work.
Overall, the results provide robust evidence for a universal Lyapunov exponent governing the chaotic dynamics of the glasma, independent of the momentum structure of the initial perturbation.

\begin{figure}[tb!]
\centering
\includegraphics[width=0.47\textwidth]{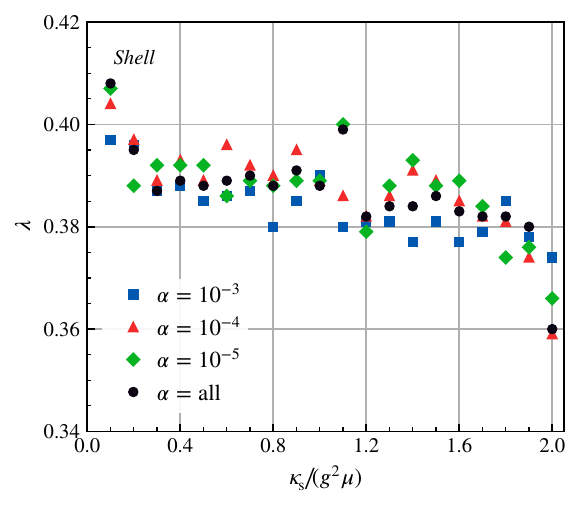}
\caption{Extracted growth rates $\lambda$ using electric field fluctuations for different initial momentum $\kappa_\mathrm{s}/(g^2\mu)$ in the shell filtering from Eq.~\eqref{Eq:Shell_Filter}, where distinct colored marker styles corresponding to different values of $\alpha$, while the black circle marker represent the combined result. The quality of the fit for $\alpha=10^{-4}$ is shown in Fig.~\ref{Fig:shelltimedep} and is similar for other values of $\alpha$.
}
\label{Fig: Shell_Filter_b_vs_k}
\end{figure}

\begin{figure}[tb!]
\centering
\includegraphics[width=0.47\textwidth]{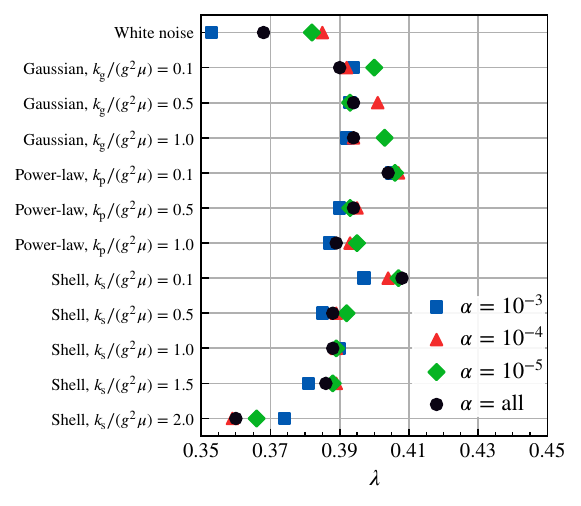}
\caption{Selection of Lyapunov exponents $\lambda$ from electric field fluctuations with different momentum space filters, with colored markers corresponding to different $\alpha$ values, together with the combined fit of these $\alpha$ values shown in a black circle marker. It is based on this compilation of results that we estimate by eye that the growth rate clusters within a band of width $0.02$ around $\lambda=0.39$.
}
\label{Fig: All_Filters_b_vs_m_OR_k}
\end{figure}

\begin{figure*}[tb!]
\includegraphics[width=0.47\textwidth]{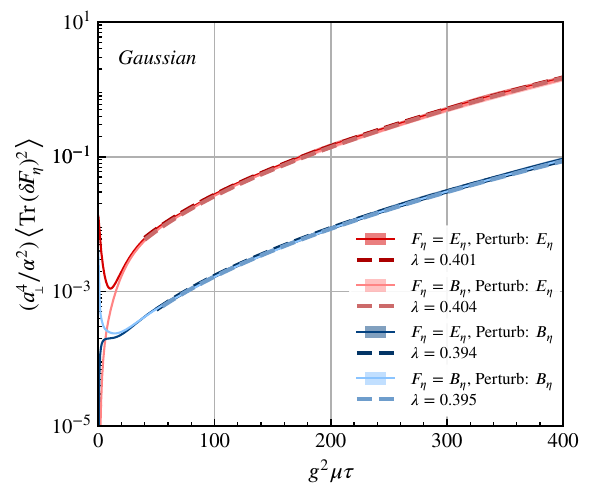}
\hfill
\includegraphics[width=0.465\textwidth]{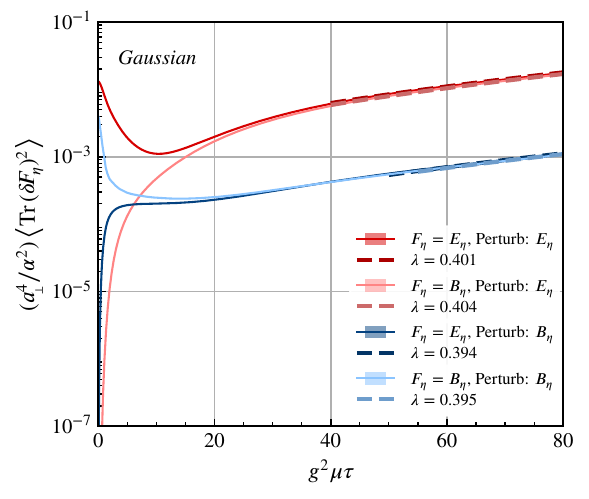}
\caption{
The electric and magnetic field perturbations $\langle \mathrm{Tr}(\delta F_\eta)^2\rangle$ as a function of time $g^2\mu\tau$, where fluctuations are initialized using a Gaussian noise with $\kappa_\mathrm{g}/(g^2\mu) = 0.5$. Here,  $\alpha =  10^{-4}$ for   perturbations in $E_\eta$  and  $\beta =  10^{-4}$ for $B_\eta$. The lines colored in red shades correspond to initially perturbing the electric field $E_\eta$ and the blue shades to an initial perturbation in $B_\eta$. The darker shades correspond to measured perturbations in the electric fields and the lighter shades to the magnetic field. Fit results are shown with dashed lines, and the corresponding value of $\lambda$ is included in the legend. The right panel zooms in on the early-time behavior of the same data. 
}
\label{Fig:Different_field_perturbations}

\end{figure*}

As is clearly shown by the parallel trajectories in Fig.~\ref{Fig:shelltimedep},  fluctuations initialized with different $\kappa_\mathrm{s}$ modes  exhibit  a remarkable universality in their growth rates.  We summarize the extracted Lyapunov exponents as a function of the shell momentum scale $\kappa_\mathrm{s}/(g^2\mu)$ in Fig.~\ref{Fig: Shell_Filter_b_vs_k}.
For each momentum shell, the growth rate $\lambda$ is obtained from an exponential
fit to the ensemble-averaged perturbation norm during the linear growth regime.
We repeat the analysis for three different perturbation amplitudes,
$\alpha \in\{ 10^{-3}, 10^{-4}, 10^{-5}\}$ as well as for a combined dataset that includes all values of $\alpha$.
The combined $\alpha=\mathrm{all}$ dataset, which benefits from the largest effective initial input, yields the most precise and stable determination of $\lambda$ across all momenta. The momentum shells span the range $\kappa_\mathrm{s}/(g^2\mu) \in [0.1, 2.0]$, sampled using twenty uniformly spaced values. This allows for a detailed probe of the possible scale dependence of the chaotic growth rate across infrared and ultraviolet modes. The important feature of Fig.~\ref{Fig: Shell_Filter_b_vs_k} is the striking consistency of the extracted Lyapunov exponent $\lambda$ within statistical uncertainties across the entire momentum range and for all perturbation amplitudes. Visible for the largest $\kappa_\mathrm{s}/(g^2\mu)$ in Fig.~\ref{Fig: Shell_Filter_b_vs_k} there is a small decrease in the values of $\lambda$. In simulations on a $N_\perp=128$ lattice, with a larger $g^2\mu a_\perp$ (not shown here), this effect is more pronounced.  This leads us to interpret this small decrease as being a result of the largest $\kappa_\mathrm{s}$ values being affected by the lattice UV cutoff rather than a physical effect. Overall, we thus conclude that we see no indications of the exponent being dependent on  $\kappa_\mathrm{s}$ within the momentum range reliably accessible on our lattices.

To provide a definitive summary of the independence of the chaotic growth rate on the momentum scale of the initial fluctuations, we present a consolidated comparison of the Lyapunov exponents extracted from all three filtering methods in  Fig.~\ref{Fig: All_Filters_b_vs_m_OR_k}. This figure summarizes the results for Gaussian, power-law, and shell filters, corresponding to the dimensionless parameters  $\kappa/(g^2\mu) \in [0.1, 2.0]$, where  $\kappa \in \{ \kappa_\mathrm{g}, \kappa_\mathrm{p}, \kappa_\mathrm{s} \}$. For each filter case, we show the exponent obtained from four different analyses: three using perturbations with amplitudes $\alpha \in\{ 10^{-3}, 10^{-4}, 10^{-5}\}$, and one using the combined $\alpha=\mathrm{all}$ dataset. 
The central and most significant result is the pronounced clustering of all data points within a band centered around $\lambda = 0.39$. This clearly demonstrates that the growth rate is insensitive to the specific functional form of the initial perturbation in momentum space. We do not consider it very meaningful to perform a fit to this collection of data, since the result would be biased by our selection of which datasets to include. Considering the totality of our results presented so far, we assign to this value an estimated systematic uncertainty of $\pm 0.02$. We emphasize that this should not be taken as a statistically rigorous error. 

We conclude that the Lyapunov exponent in the glasma has an averaged value of
\begin{equation}
    \lambda = 0.39 \pm 0.02,
\end{equation}
consistent across multiple momentum scales of the initial fluctuations.

\subsection{Electric versus magnetic field perturbations}
\label{subsec:elmagpert}

Having established the scale independence of the Lyapunov exponent, we now analyze how perturbations introduced in different field components propagate across the dynamical degrees of freedom to further test the robustness and universality of the chaotic growth. To this end, we investigate whether the chaotic growth exhibits any dependence on which specific field component is perturbed or measured. We use here a Gaussian spatial filter with characteristic mass scale $\kappa_\mathrm{g}/(g^2\mu) = 0.5$, initialize the system with a perturbation in only one field sector (either $E_\eta$ or $B_\eta$), and monitor the subsequent growth in both components. Figure~\ref{Fig:Different_field_perturbations} presents the results of this analysis.

Despite different initial conditions, all four perturbation measurement combinations exhibit a clear exponential growth regime following a short transient period. Remarkably, the extracted Lyapunov exponents are found to be nearly identical across all cases, with values $\lambda \approx 0.40$
within statistical uncertainties. These are consistent with the values extracted from purely electric-field fluctuations identified in previous scale-dependence studies. Furthermore, we have verified that the growth rate remains robust across a wider range of configurations; supplementary simulations with white-noise perturbations and different types of filters for various parameter sets consistently yield similar results, though these are not presented here for brevity. We conclude that $\lambda$  is insensitive to whether the perturbation is introduced in the electric or magnetic sector, and likewise independent of which field component is used to quantify the perturbation growth.

While the exponential growth rates are nearly identical, the initial amplitudes differ between the four cases.
The zoom-in at early times shown in the right panel of Fig.~\ref{Fig:Different_field_perturbations} shows more clearly how each type of perturbation is initially present only in the mode in which it was initialized. 
When the perturbation is applied to $E_\eta$, the short-time response in $\langle\mathrm{Tr}(\delta B_\eta)^2\rangle$ is significantly slower than the response in $\langle\mathrm{Tr}(\delta E_\eta)^2\rangle$ in the opposite case. Nevertheless, these differences do not affect the slope of the exponential growth, as seen on the logarithmic scale. This result shows that the unstable mode couples the two polarization states of the glasma fields, which, in a linearized approximation, are independent of each other.

\begin{figure}[tb!]
		\centering
		\includegraphics[width=\linewidth]{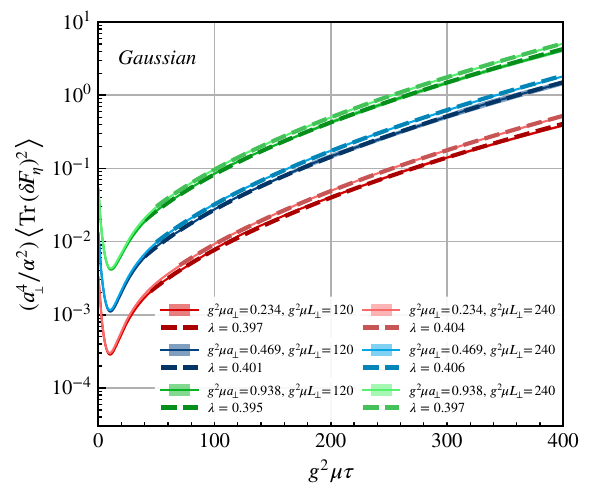}   
\caption{Dependence of the rescaled $\langle \mathrm{Tr}(\delta E_\eta)^2\rangle$  on  the lattice volume $g^2\mu L_\perp$. The darker and the lighter shades show
$g^2\mu L_\perp=120$ and  $g^2\mu L_\perp=240$ respectively, with different colors corresponding to  different $g^2\mu a_\perp$.
}
\label{Fig: L_dependence}
\end{figure}

\begin{figure}[tb!]
	\centering
\includegraphics[width=\linewidth]{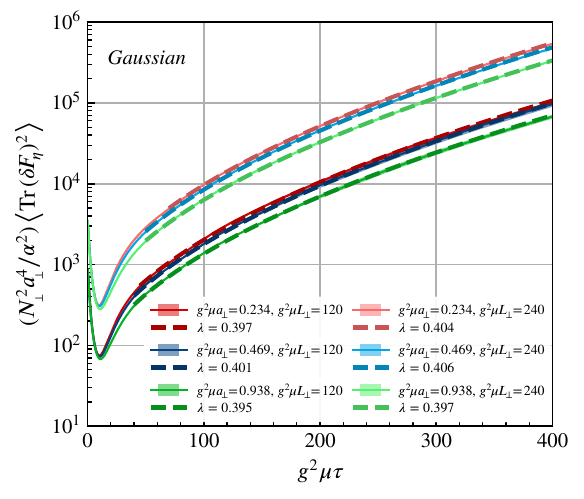}
\caption{ Dependence of the rescaled $\langle \mathrm{Tr}(\delta E_\eta)^2\rangle$ on the lattice spacing $g^2\mu a_\perp$. Different colors correspond to different lattice spacings.
Note that the data in this plot is the same as in Fig.~\ref{Fig: L_dependence}, but scaled by $N_\perp^2$ so that the data corresponding to different  $g^2\mu a_\perp$ coincide at $g^2\mu \tau=0$. 
\label{Fig: a_dependence}
}
\end{figure}

\subsection{Lattice parameter dependence}
\label{subsec:lattice_dependence}
Having established the universality of the Lyapunov exponent across different perturbation scales, filter types, and field components, we now investigate its dependence on the underlying lattice discretization parameters. This is a crucial systematic check to ensure that the extracted chaotic growth rate represents a genuine physical property of the glasma rather than an artifact of the numerical implementation. To this end, we perform a dedicated study by varying both the physical transverse lattice volume $g^2\mu L_\perp$ and the lattice spacing $g^2\mu a_\perp$. We employ a Gaussian-filtered noise with $\kappa_\mathrm{g}/(g^2\mu)=0.5$ and fix the perturbation amplitude to $\alpha = 10^{-4}$. Here we show results for the evolution of the perturbed electric fields $\langle \Tr (\delta E_\eta)^2\rangle$, but we have also performed similar checks for the magnetic field. 

As derived from the noise construction in Eq.~\eqref{Eq:Eeta_growth_scaling}, at the initial time $\tau=0$, the perturbation measure scales as $a_\perp^4 \langle \mathrm{Tr}(\delta E_\eta)^2\rangle \big|_{\tau=0} \propto (\kappa  a_\perp)^2$. This scaling informs our strategy for isolating dependencies as follows: To examine the dependence on the physical volume $L_\perp$ at fixed lattice spacing, we extract $a_\perp^4 \langle \mathrm{Tr}(\delta E_\eta)^2\rangle$ directly. According to the scaling, at $\tau=0$ this quantity is independent of $L_\perp$ (for a fixed $a_\perp$), providing a common starting point. Next, to examine the dependence on the lattice spacing $a_\perp$ at fixed physical volume, we evaluate the rescaled quantity $N_\perp^2 a_\perp^4 \langle \mathrm{Tr}(\delta E_\eta)^2\rangle$, such that $N_\perp^2 a_\perp^4 \langle \mathrm{Tr}(\delta E_\eta)^2\rangle \big|_{\tau=0} \propto (\kappa  L_\perp)^2$. At $\tau=0$, this rescaled measure is independent of $a_\perp$ and depends on $L_\perp$, which makes the comparison easier.

Figure~\ref{Fig: L_dependence} presents the volume dependence. We show results for three lattice spacings $g^2\mu a_\perp\in\{0.234,\,0.469,\,0.938\}$, each for two physical volumes $g^2\mu L_\perp\in\{120, 240\}$. At early times, the curves for the same $g^2\mu a_\perp$ but different $g^2\mu L_\perp$ overlap almost perfectly, as expected from the scaling. This shows that the normalization is not dominated by finite-volume effects and that the onset of the unstable growth is insensitive to the infrared cutoff. At later times, a mild separation between the two values of $g^2\mu L_\perp$ becomes visible, indicating a small finite-volume effect in the overall amplitude. Importantly, however, the late-time growth rate extracted from the exponential fits remains essentially unchanged when increasing the volume. This clearly demonstrates that the extracted Lyapunov exponent is insensitive to the finite volume of our simulation, provided that the physical volume is sufficiently large to contain the relevant unstable modes. 

The volume $L_\perp$ acts as an infrared cutoff (see discussion of Eq.~\eqref{Eq: Poisson_Eq_solution_momentum_space_2} above), and the insensitivity of the growth rate to $g^2\mu L_\perp$ can be interpreted as an independence of the infrared cutoff provided by $L_\perp$. On the other hand, the numerical calculation only depends on the dimensionless combinations of the parameters. Only after we assign a value in physical units to $g^2\mu$ or $L_\perp$, we have a value in physical units for the other. Thus, the insensitivity of $\lambda$ to $g^2\mu L_\perp$ can also be seen as an independence on $g^2\mu$ for a fixed lattice size $L_\perp$, indicating that the functional form  $ \exp(\lambda\sqrt{g^2 \mu \tau})$ is the correct dependence on the saturation scale.

Next, we study the dependence on the lattice spacing $a_\perp$. Here, we rescale the same data by an explicit factor of $N_\perp^2$ such that the value at $\tau=0$ is independent of $g^2\mu a_\perp$ and instead depends only on the physical volume. The corresponding comparison is shown in Fig.~\ref{Fig: a_dependence}.
We observe that the rescaled growth trajectories for different $g^2\mu a_\perp$ (at fixed $g^2\mu L_\perp = 120$ or $240$) lie remarkably close to each other. More importantly, at late times, the extracted Lyapunov exponents are stable across a variation in lattice spacing by a factor of four. The fitted values, noted in the figure, cluster tightly around $\lambda\approx 0.40$, with variations at the level of a few percent. This confirms that the Lyapunov exponent is ultraviolet convergent and essentially insensitive to the UV cutoff within our numerical accuracy.

Taken together, Figs.~\ref{Fig: L_dependence} and~\ref{Fig: a_dependence} show that the chaotic growth rate in the glasma is largely insensitive to changes in both the infrared regulator (volume $L_\perp$) and the ultraviolet regulator (lattice spacing $a_\perp$) in the parameter ranges explored here. Thus, the maximal Lyapunov exponent of the glasma is independent of the numerical discretization details.

\section{Summary and conclusions}
\label{Section: Conclusions}

 Many aspects of the thermalization process in the early stages of a heavy-ion collision are still poorly understood. In this paper, we have focused on the chaotic behavior in the boost-invariant expanding two-dimensional gauge field system, the glasma. We have done this in a very straightforward way by initializing small perturbations in the fields and following their time evolution with the equations of motion. The presence of classical chaotic behavior is, in this case, signaled by exponentially growing modes. By following the system until late enough time, where only the leading most unstable mode dominates, we can extract the corresponding Lyapunov exponent from its growth rate. 

 We perform the numerical analysis within the standard boost-invariant numerical setup for studying glasma fields. Staying in a regime where the perturbations are linear, which we have verified by varying their initial amplitudes, we have first discovered that their time dependence at late times can be very well described by a simple functional form $\sim \exp ( \lambda\sqrt{g^2\mu \tau}).$ To extract the value of the parameter $\lambda$, we then developed a systematic fitting procedure in order to find a time window that is long enough to provide a good sensitivity to the value of $\lambda$ but nevertheless limited to a range where the functional form works. To achieve this, we perform fits over different time intervals, taking into account the statistical errors in our data. We  minimize the parameter uncertainty $\delta\lambda$ from the fit over the set of candidate intervals $[t_{\min}, t_{\max}]$ that satisfy $\chi^2_{\mathrm{red}} < 1$. Having tested and developed this robust fitting procedure, we then use it systematically for different perturbation types (electric and magnetic) and for various parameter sets, allowing for systematic comparison of the extracted Lyapunov exponents.

It would seem natural that a collective mode such as the one we are looking for here would be dominated by relatively low-momentum degrees of freedom, in particular since its growth rate is parametrically similar to the plasmon or Debye mass scale in the system. To get some access to the momentum structure of the unstable mode, we initialized fluctuations with different momentum modes, using  Gaussian, power-law, and broadened delta function (momentum shell) functional forms. Within these comparisons, we find that the late time growth rate of the unstable mode is unaffected, but the amplitude is lower if the perturbations are initialized at larger $k_\perp$. 
From this, we can see that the unstable mode couples to all momenta in the initial condition, but has a larger overlap with more infrared degrees of freedom. The glasma fields, if linearized, have two independent polarization states ($E_\perp \& B_\eta$ and $E_\eta \& B_\perp$). By exciting and measuring them separately, we have seen that the same unstable mode couples to both polarization states. We have also tested that the growth rate is (for a fixed saturation scale $g^2\mu$) independent of discretization parameters: the lattice spacing $a_\perp$ and the physical volume of the system $L_\perp$. 

Combining our numerical simulation results, we find the time dependence of the unstable mode to be $\delta F (\tau) \sim \exp (\lambda/2 \sqrt{g^2\mu \tau}),$ where $\delta F_\eta=\delta E_\eta, \delta B_\eta$ with $\lambda \approx 0.39\pm 0.02$. Here, our uncertainty estimate is dominated by systematics, i.e., varying the initial conditions for the fluctuations and lattice parameters, rather than statistics. 
Expressed in terms of a time-dependent growth rate  
\begin{equation}
\frac{\ud}{\ud \tau}\delta F(\tau) =  \gamma(\tau) \delta F(\tau),
\end{equation}
this would translate into a value 
\begin{equation}
    \gamma(\tau) = (0.1\pm 0.005 )\sqrt{\frac{g^2 \mu}{\tau}}.
    \label{eq:growthrate}
\end{equation}
We emphasize that this number is remarkably robust and stays within this narrow range across variations in the initial conditions for the fluctuations.

The first study presented here points towards several avenues for future research. As a relatively straightforward step, we plan to extend the simulations to the SU($3$) group, and to investigate the effect of regulating the infrared in the Poisson Eq.~\eqref{Eq: Poisson_Eq_solution_momentum_space_2} by a mass term rather than the lattice volume. We do not expect major qualitative changes from these, but it will be interesting to extract the  $\nc$-dependence of the growth rate. An important, more detailed study would be to analyze the momentum structure of the unstable mode in more detail, by Fourier decomposing it at a late stage of the evolution. Such a momentum spectrum will require us to address the gauge-dependence, e.g., by imposing a Coulomb gauge condition at the time of measurement. Here, we have also just studied the leading Lyapunov exponent from the late-time behavior. Much more insight could be gained by diagonalizing the whole Hamiltonian in order to find the full spectrum of eigenvalues, as done in \cite{Kunihiro:2010tg}. It remains to be seen whether this would be possible on equally large lattices as we are using here.

The growth rate of the unstable mode, as expressed in Eq.~\eqref{eq:growthrate}, is parametrically similar to the expected behavior of the plasmon or Debye mass scales in the boost invariant expanding glasma~\cite{Krasnitz:2000gz, Lappi:2017ckt}. It would be interesting to understand in more detail the relation of the unstable mode we have discovered here to those scales. 
When the boost invariance of the glasma is broken by small fluctuations, it will exhibit a different kind of exponentially growing modes related to the Weibel instability~\cite{Romatschke:2005pm, Romatschke:2006nk, Berges:2011sb, Schlichting:2012es, Berges:2012cj}. The relation of the boost-invariant exponential growth to this instability, which provides the first step in the isotropization process, is not yet understood. Another interesting topic for the future would be to look into chaotic behavior for quarks interacting with the glasma fields. By using the Wong's equation approach~\cite{Avramescu:2023qvv, Avramescu:2024poa, Avramescu:2024xts, Ruggieri:2018rzi, Liu:2020cpj,  Khowal:2021zoo, Ruggieri:2022kxv, Pooja:2022ojj, Pooja:2023gqt,  Pooja:2024rnn, Das:2022lqh, Oliva:2024rex, Parisi:2025slf} one could extract the Lyapunov exponent directly from the particle trajectories. We plan to return to these topics in future work.

\section*{Acknowledgment}
This work was supported by the Research Council of Finland, the Centre of Excellence in Quark Matter (project 346324 and 364191), and by the European Research Council (ERC, grant agreement No. ERC-2018-ADG835105 YoctoLHC). 
DA also acknowledges the support of the Vilho, Yrj\"{o} and Kalle V\"{a}is\"{a}l\"{a} Foundation. 
The content of this article does not reflect the official opinion of the European Union and responsibility for the information and views expressed therein lies entirely with the
authors. 
Computing resources from CSC – IT Center for Science in Espoo, Finland, and the Finnish Grid and Cloud Infrastructure (persistent identifier \texttt{urn:nbn:fi:research-infras-2016072533}) were used in this work.

\clearpage

\begin{widetext}
\appendix

\section{Fit results}
\label{Appendix:Fit_Results}

This Appendix presents the results of the fits to different datasets,
for the white noise perturbations in Table \ref{Table: NoFilter}, the Gaussian perturbations in Table~\ref{Table: Gaussian_Filter}, the power law perturbations in  Table~\ref{Table: Powerlaw_Filter} and for the momentum shell filtering in 	Table~\ref{Table: Shell_Filter}. For the latter, we show a selection from our full dataset.

\begin{table*}[h!]
\caption{Fit parameters for perturbed $E_\eta$ field evolution without filtering. }
\centering
\begin{ruledtabular}
\begin{tabular}{@{}ccccc@{}}
			$\alpha$ & Fit Range & $a~(\times 10^{-3})$ & $\lambda$ & $\chi^2_{\mathrm{red}}$ \\
			\hline\hline			
    $10^{-3}$ & [65,230] & 3.079 & 0.353 & 0.927 \\
    $10^{-4}$ & [70,400] & 2.259 & 0.385 & 0.395 \\
    $10^{-5}$ & [70,395] & 2.335 & 0.382 & 0.996 \\
    all       & [70,170] & 2.671 & 0.368 & 0.994 \\
\end{tabular}
\end{ruledtabular}
\label{Table: NoFilter}
\end{table*}

\begin{table*}[h!]
\caption{Fit parameters for perturbed $E_\eta$ field evolution with Gaussian filtering.}
\centering
\begin{ruledtabular}
\begin{tabular}{@{}cccccc@{}}
$\kappa_\mathrm{g}/(g^2\mu)$ & $\alpha$ & Fit Range & $a~(\times 10^{-4})$ & $\lambda$ & $\chi^2_{\mathrm{red}}$ \\
\hline\hline
    0.1 & $10^{-3}$ & [80,400] & 0.580 & 0.394 & 0.922 \\
    0.1 & $10^{-4}$ & [55,400] & 0.560 & 0.392 & 0.936 \\
    0.1 & $10^{-5}$ & [45,400] & 0.510 & 0.400 & 0.653 \\
    0.1 & all       & [85,400] & 0.600 & 0.390 & 0.978 \\
    \\    
    0.5 & $10^{-3}$ & [45,245] & 5.280 & 0.393 & 0.983 \\
    0.5 & $10^{-4}$ & [40,400] & 4.960 & 0.401 & 0.322 \\
    0.5 & $10^{-5}$ & [40,400] & 5.200 & 0.393 & 0.862 \\
    0.5 & all       & [45,255] & 5.250 & 0.394 & 0.986 \\
    \\
    1.0 & $10^{-3}$ & [40,200] & 9.830 & 0.392 & 0.969 \\
    1.0 & $10^{-4}$ & [35,400] & 9.560 & 0.394 & 0.962 \\
    1.0 & $10^{-5}$ & [35,400] & 9.030 & 0.403 & 0.947 \\
    1.0 & all       & [55,195] & 9.630 & 0.394 & 0.995 \\            
			
\end{tabular}
\end{ruledtabular}
\label{Table: Gaussian_Filter}
\end{table*}

\begin{table*}[h!]
\caption{Fit parameters for perturbed $E_\eta$ field evolution with power-law filtering. 
}
\centering
\begin{ruledtabular}
\begin{tabular}{@{}cccccc@{}}
$\kappa_\mathrm{p}/(g^2\mu)$ & $\alpha$ & Fit Range & $a~(\times 10^{-3})$ & $\lambda$ & $\chi^2_{\mathrm{red}}$ \\
\hline\hline
0.1 & $10^{-3}$ & [50,290] & 0.079 & 0.405 & 0.998 \\
0.1 & $10^{-4}$ & [40,400] & 0.075 & 0.407 & 0.918 \\
0.1 & $10^{-5}$ & [40,400] & 0.076 & 0.406 & 0.888 \\
0.1 & all       & [45,320] & 0.078 & 0.404 & 0.966 \\
			\\
0.5 & $10^{-3}$ & [45,205] & 0.650 & 0.390 & 0.955 \\
0.5 & $10^{-4}$ & [40,400] & 0.619 & 0.395 & 0.635 \\
0.5 & $10^{-5}$ & [40,400] & 0.628 & 0.393 & 0.771 \\
0.5 & all       & [45,205] & 0.630 & 0.394 & 0.930 \\
			\\
1.0 & $10^{-3}$ & [40,200] & 1.133 & 0.387 & 0.924 \\
1.0 & $10^{-4}$ & [35,400] & 1.074 & 0.393 & 0.778 \\
1.0 & $10^{-5}$ & [35,400] & 1.061 & 0.395 & 0.948 \\
1.0 & all       & [45,200] & 1.117 & 0.389 & 0.997 \\
\end{tabular}
\end{ruledtabular}
\label{Table: Powerlaw_Filter}
\end{table*}

\begin{table*}[h!]
	\caption{A selection of fit parameters for perturbed $E_\eta$ field evolution with shell filtering.
    }
	\centering
	\begin{ruledtabular}
		\begin{tabular}{@{}cccccc@{}}
			$\kappa_\mathrm{s}/(g^2\mu)$ & $\alpha$ & Fit Range & $a~(\times 10^{-4})$ & $\lambda$ & $\chi^2_{\mathrm{red}}$ \\
			\hline\hline
0.1 & $10^{-3}$ & [50,260] & 1.920 & 0.397 & 0.977 \\
0.1 & $10^{-4}$ & [40,400] & 1.780 & 0.404 & 0.878 \\
0.1 & $10^{-5}$ & [45,400] & 1.800 & 0.407 & 0.454 \\
0.1 & all       & [45,230] & 1.760 & 0.408 & 0.984 \\
			\\
0.5 & $10^{-3}$ & [40,345] & 1.650 & 0.385 & 0.964 \\
0.5 & $10^{-4}$ & [35,400] & 1.580 & 0.389 & 0.575 \\
0.5 & $10^{-5}$ & [35,400] & 1.530 & 0.392 & 0.840 \\
0.5 & all       & [40,330] & 1.600 & 0.388 & 0.998 \\
			\\
1.0 & $10^{-3}$ & [60,360] & 0.940 & 0.390 & 0.988 \\
1.0 & $10^{-4}$ & [50,400] & 0.960 & 0.389 & 0.784 \\
1.0 & $10^{-5}$ & [55,400] & 0.940 & 0.389 & 0.687 \\
1.0 & all       & [60,400] & 0.950 & 0.388 & 0.940 \\
			\\
1.5 & $10^{-3}$ & [65,400] & 0.430 & 0.381 & 0.556 \\
1.5 & $10^{-4}$ & [75,400] & 0.380 & 0.389 & 0.812 \\
1.5 & $10^{-5}$ & [70,400] & 0.400 & 0.388 & 0.387 \\
1.5 & all       & [70,400] & 0.400 & 0.386 & 0.791 \\
			\\
2.0 & $10^{-3}$ & [100,380] & 0.170 & 0.374 & 0.998 \\
2.0 & $10^{-4}$ & [90,260] & 0.200 & 0.359 & 0.991 \\
2.0 & $10^{-5}$ & [100,300] & 0.180 & 0.366 & 0.993 \\
2.0 & all       & [90,260] & 0.200 & 0.360 & 0.971 \\
		\end{tabular}
	\end{ruledtabular}
	\label{Table: Shell_Filter}
\end{table*}

\FloatBarrier
\end{widetext}

\bibliographystyle{JHEP}
\bibliography{Lyapunov_PRD_v2_refs}

\end{document}